\documentclass[aps,preprint]{revtex4}%
\usepackage{amsfonts}
\usepackage{amsmath}
\usepackage{amssymb}
\usepackage{subfigure}
\usepackage{graphicx}
\usepackage{array}%
\begin{document}
\preprint{CTP-SCU/2019001}
\title{Thermodynamics and Phase Transition of a Nonlinear Electrodynamics Black Hole
in a Cavity}
\author{Peng Wang}
\email{pengw@scu.edu.cn}
\author{Houwen Wu}
\email{iverwu@scu.edu.cn}
\author{Haitang Yang}
\email{hyanga@scu.edu.cn}
\affiliation{Center for Theoretical Physics, College of Physical Science and Technology,
Sichuan University, Chengdu, 610064, China}

\begin{abstract}
We discuss the thermodynamics of a general nonlinear electrodynamics (NLED)
asymptotically flat black hole enclosed in a finite spherical cavity. A
canonical ensemble is considered, which means that the temperature and the
charge on the wall of the cavity are fixed. After the free energy is obtained
by computing the Euclidean action, it shows that the first law of
thermodynamics is satisfied at the locally stationary points of the free
energy. Focusing on a Born-Infeld (BI) black hole in a cavity, the phase
structure and transition in various regions of the parameter space are
investigated. In the region where the BI electrodynamics has weak
nonlinearities, Hawking-Page-like and van der Waals-like phase transitions
occur, and a tricritical point appears. In the region where the BI
electrodynamics has strong enough nonlinearities, only Hawking-Page-like phase
transitions occur. The phase diagram of the BI black hole in a cavity can have
dissimilarity from that of a BI black hole using asymptotically anti-de Sitter
boundary conditions. The dissimilarity may stem from a lack of an appropriate
reference state with the same charge and temperature for the BI-AdS black hole.

\end{abstract}
\maketitle
\tableofcontents

\bigskip



\section{Introduction}

A Schwarzschild black hole in asymptotically flat space has negative specific
heat and hence radiates more when it is smaller. To make this system thermally
stable, appropriate boundary conditions must be imposed. One popular choice is
putting the black hole in anti-de Sitter (AdS) space, which has a negative
cosmological constant. The black hole becomes thermally stable since the AdS
boundary acts as a reflecting wall. The thermodynamic properties of AdS black
holes were first studied by Hawking and Page \cite{IN-Hawking:1982dh}, who
discovered the Hawking-Page phase transition, i.e., a phase transition between
the thermal AdS space and the Schwarzschild-AdS black hole. Later, with the
advent of the AdS/CFT correspondence
\cite{IN-Maldacena:1997re,IN-Gubser:1998bc,IN-Witten:1998qj}, there has been
much interest in studying the phase transitions of AdS black holes
\cite{IN-Witten:1998zw,IN-Chamblin:1999tk,IN-Chamblin:1999hg,IN-Caldarelli:1999xj,IN-Cai:2001dz,IN-Kubiznak:2012wp}%
. However, it is not clear whether the duality between the black hole and a
boundary field theory is independent of the details of the boundary
conditions, or just the special result of the asymptotically AdS space. It is
therefore interesting to investigate thermodynamics of black holes in the case
of different boundary conditions.

Alternatively, one can place the black hole inside a cavity in asymptotically
flat space, on the wall of which the metric is fixed. In \cite{IN-York:1986it}%
, York showed that a Schwarzschild black hole in a cavity can be thermally
stable and experiences a Hawking-Page-like transition to the thermal flat
space as the temperature decreases. Later, the thermodynamics of a
Reissner-Nordstrom (RN) black hole in a cavity was discussed in a grand
canonical ensemble \cite{IN-Braden:1990hw} and a canonical ensemble
\cite{IN-Carlip:2003ne,IN-Lundgren:2006kt}. Similar to a RN-AdS black hole, it
was found that a Hawking-Page-like phase transition occurs in the grand
canonical ensemble, and a van der Waals-like phase transition occurs in the
canonical ensemble. Note that a van der Waals-like phase transition consists
of a first-order phase transition between two black hole phases of different
sizes and a critical point, where the first-order phase transition ends, and a
second order phase transition takes place. The phase structures of several
black brane systems in a cavity were investigated in a series of paper
\cite{IN-Lu:2010xt,IN-Wu:2011yu,IN-Lu:2012rm,IN-Lu:2013nt,IN-Zhou:2015yxa,IN-Xiao:2015bha}%
, where Hawking-Page-like or van der Waals-like phase transitions were always
found except for some special cases. Including charged scalars, boson stars
and hairy black holes in a cavity were considered in
\cite{IN-Basu:2016srp,IN-Peng:2017gss,IN-Peng:2017squ,IN-Peng:2018abh}, which
showed that the phase structure of the gravity system in a cavity is
strikingly similar to that of holographic superconductors in the AdS gravity.
The stabilities of solitons, stars and black holes in a cavity were also
studied in
\cite{IN-Sanchis-Gual:2015lje,IN-Dolan:2015dha,IN-Ponglertsakul:2016wae,IN-Sanchis-Gual:2016tcm,IN-Ponglertsakul:2016anb,IN-Sanchis-Gual:2016ros,IN-Dias:2018zjg,IN-Dias:2018yey}%
. It was found that the nonlinear dynamical evolution of a charged black hole
in a cavity could end in a quasi-local hairy black hole. Recently, McGough,
Mezei and Verlinde \cite{IN-McGough:2016lol} proposed that the $T\bar{T}$
deformed CFT$_{\text{2}}$ locates at the finite radial position of
AdS$_{\text{3}}$, which further motivates us to explore the properties of a
black hole in a cavity.

Taking quantum contributions into account, nonlinear corrections are usually
added to the Maxwell Lagrangian, which gives the nonlinear electrodynamics
(NLED). Coupling NLED fields to gravity, various NLED charged black holes were
derived and discussed in a number of papers
\cite{IN-Soleng:1995kn,IN-Maeda:2008ha,IN-Hendi:2017mgb,IN-Tao:2017fsy,IN-Guo:2017bru,IN-Mu:2017usw,IN-Hendi:2012um,IN-Mo:2016jqd,IN-Nam:2018tpf,IN-Dehghani:2018eps}%
. It is interesting to note that some NLED black holes can be regular black
hole models \cite{IN-AyonBeato:1998ub,IN-AyonBeato:1999rg}. As pointed out in
\cite{IN-Bronnikov:2000vy}, a globally regular NLED black hole requires
vanishing electric charge and a finite NLED Lagrangian (or in the FP dual
theory). Born-Infeld (BI) electrodynamics was first introduced to incorporates
maximal electric fields and smooths divergences of the electrostatic
self-energy of point charges \cite{IN-Born:1934gh}. Later, it is realized that
BI electrodynamics can come from the low energy limit of string theory and
encodes the low-energy dynamics of D-branes. The BI black hole solution was
obtained in \cite{IN-Dey:2004yt,IN-Cai:2004eh}. The thermodynamic behavior and
phase transitions of BI black holes in various gravities were investigated in
\cite{IN-Fernando:2003tz,IN-Fernando:2006gh,IN-Banerjee:2010da,IN-Banerjee:2011cz,IN-Lala:2011np,IN-Banerjee:2012zm,IN-Azreg-Ainou:2014twa,IN-Hendi:2015hoa,IN-Zangeneh:2016fhy,IN-Zeng:2016sei,IN-Li:2016nll,IN-Zou:2013owa,IN-Hendi:2017oka}%
. Especially, the thermodynamics of a 4D BI-AdS black hole was studied in
\cite{IN-Gunasekaran:2012dq,IN-Dehyadegari:2017hvd,IN-Wang:2018xdz}, where a
reentrant phase transition was always observed in a certain region of the
parameter space.

In this paper, we first investigate the thermodynamic behavior of\ a 4D
general NLED asymptotically flat black hole enclosed in a cavity. Then, we
turn to study the phase structure and transition of a BI black hole in a
cavity. We find that Hawking-Page-like and van der Waals-like phase
transitions can occur while there is no reentrant phase transition. The rest
of this paper is organized as follows. In section \ref{Sec:NBHC}, we compute
the Euclidean action for the general NLED black hole in a cavity and discuss
the thermodynamic properties of the system in the canonical ensemble. In
section \ref{Sec:BIBHC}, we focus on the BI black hole case to discuss the
phase structure and transition. The phase diagrams of the BI black hole in a
cavity is given in FIG. \ref{fig:RP}, from which one can read the black hole's
phase structure and transition. In the appendix, we present an alternative
derivation of the Euclidean action for a general NLED black hole in a cavity
using the reduced action method proposed in \cite{IN-Braden:1990hw}.

\section{NLED Black Hole in a Cavity}

\label{Sec:NBHC}

In this section, we consider a NLED charged black hole inside a cavity, on the
boundary of which the temperature and charge are fixed. That said, the
thermodynamics of the black hole is discussed in a canonical ensemble.

\subsection{Black Hole Solution}

First, we will consider the black hole solution in a $\left(  3+1\right)  $
dimensional model of gravity coupled to a nonlinear electromagnetic field
$A_{\mu}$. On a spacetime manifold $\mathcal{M}$ with a time-like boundary
$\partial\mathcal{M}$, the action is given by%
\begin{equation}
\mathcal{S}=\int_{\mathcal{M}}d^{4}x\sqrt{-g}\left[  R+\mathcal{L}\left(
s,p\right)  \right]  +\mathcal{S}_{\text{surf}}, \label{eq:Action}%
\end{equation}
where we take $16\pi G=1$ for simplicity, $\mathcal{L}\left(  s,p\right)  $ is
a general NLED Lagrangian, and $S_{\text{surf}}$ are the surface terms on
$\partial\mathcal{M}$. Here, $s$ and $p$ are two independent nontrivial
scalars built from the field strength tensor $F_{\mu\nu}=\partial_{\mu}A_{\nu
}-\partial_{\nu}A_{\mu}$ and none of its derivatives:
\begin{equation}
s=-\frac{1}{4}F^{\mu\nu}F_{\mu\nu}\text{ and }p=-\frac{1}{8}\epsilon^{\mu
\nu\rho\sigma}F_{\mu\nu}F_{\rho\sigma}\text{,}%
\end{equation}
where $\epsilon^{\mu\nu\rho\sigma}\equiv-\left[  \mu\text{ }\nu\text{ }%
\rho\text{ }\sigma\right]  /\sqrt{-g}$ is a totally antisymmetric Lorentz
tensor, and $\left[  \mu\text{ }\nu\text{ }\rho\text{ }\sigma\right]  $ is the
permutation symbol. For later use, we define%
\begin{equation}
G^{\mu\nu}=-\frac{\partial\mathcal{L}\left(  s,p\right)  }{\partial F_{\mu\nu
}}=\mathcal{L}^{\left(  1,0\right)  }\left(  s,p\right)  F^{\mu\nu}+\frac
{1}{2}\mathcal{L}^{\left(  0,1\right)  }\left(  s,p\right)  \epsilon^{\mu
\nu\rho\sigma}F_{\rho\sigma}.
\end{equation}
where we denote $\mathcal{L}^{\left(  1,0\right)  }\left(  s,p\right)
\equiv\frac{\partial\mathcal{L}\left(  s,p\right)  }{\partial s}$ and
$\mathcal{L}^{\left(  0,1\right)  }\left(  s,p\right)  \equiv\frac
{\partial\mathcal{L}\left(  s,p\right)  }{\partial p}$, respectively. Note
that the general NLED theories with the Lagrangian $\mathcal{L}\left(
s,p\right)  $ were first considered in
\cite{NBHC-Pellicer:1969cf,NBHC-Pleb:1969}. The surface terms of the action
$\left(  \ref{eq:Action}\right)  $ are
\begin{equation}
\mathcal{S}_{\text{surf}}=-2\int_{\partial\mathcal{M}}d^{3}x\sqrt{-\gamma
}\left(  K-K_{0}\right)  -\int_{\partial\mathcal{M}}d^{3}x\sqrt{-\gamma}%
n_{\nu}G^{\mu\nu}A_{\mu}.
\end{equation}
The first term above is the Gibbons-Hawking-York surface term, where $K$ is
the extrinsic curvature, $\gamma$ is the metric on the boundary, and $K_{0}$
is a subtraction term to make the Gibbons-Hawking-York term vanish in flat
spacetime. When the metric on $\partial\mathcal{M}$ is fixed, the
Gibbons-Hawking-York term is crucial to obtain the correct the equations of
motion from performing the variation. The second term, where $n^{\mu}$ is the
unit outward-pointing normal vector of $\partial\mathcal{M}$, is included to
keep the charge fixed on $\partial\mathcal{M}$, instead of the potential, when
one varies the action to have the correct equations of motion
\cite{IN-Wang:2018xdz}. Varying the action $\left(  \ref{eq:Action}\right)  $
in terms of $g_{\mu\nu}$ and $A_{\mu}$ with the metric and the charge fixed on
$\partial\mathcal{M}$, we find that the equations of motion are%
\begin{align}
R_{\mu\nu}-\frac{1}{2}Rg_{\mu\nu}  &  =\frac{T_{\mu\nu}}{2}\text{,}\nonumber\\
\nabla_{\mu}G^{\mu\nu}  &  =0\text{,}%
\end{align}
where $T_{\mu\nu}$ is the energy-momentum tensor for the NLED field:%
\begin{equation}
T_{\mu\nu}=g_{\mu\nu}\left[  \mathcal{L}\left(  s,p\right)  -p\mathcal{L}%
^{\left(  0,1\right)  }\left(  s,p\right)  \right]  +\mathcal{L}^{\left(
1,0\right)  }\left(  s,p\right)  F_{\mu}^{\text{ }\rho}F_{\nu\rho}\text{.}
\label{eq:stNLED}%
\end{equation}

We consider a static spherically symmetric black hole solution with the metric
and the NLED field of the form%
\begin{align}
ds^{2}  &  =-f\left(  r\right)  dt^{2}+\frac{dr^{2}}{f\left(  r\right)
}+r^{2}\left(  d\theta^{2}+\sin^{2}\theta d\phi^{2}\right)  \text{,}%
\nonumber\\
A  &  =A_{t}\left(  r\right)  dt\text{.} \label{eq:ansatz}%
\end{align}
Moreover, we assume that the black hole lives in a spherical cavity, which has
a boundary $\partial\mathcal{M}$ at $r=r_{B}$. The spacelike slices with
constant $t$ of $\partial\mathcal{M}$ are $2$-spheres $S^{2}$ whose radii are
$r_{B}$. The equations of motion then reduce to
\begin{align}
-1+f\left(  r\right)  +rf^{\prime}\left(  r\right)   &  =\frac{r^{2}}%
{2}\left[  \mathcal{L}\left(  s,0\right)  +A_{t}^{\prime}\left(  r\right)
G^{rt}\right]  ,\label{eq:ttEOM}\\
2f^{\prime}\left(  r\right)  +rf^{\prime\prime}\left(  r\right)   &
=r\mathcal{L}\left(  s,0\right)  ,\label{eq:thetathetaEOM}\\
\left[  r^{2}G^{rt}\right]  ^{\prime}  &  =0\text{,} \label{eq:NLEDEOM}%
\end{align}
where
\begin{equation}
s=\frac{A_{t}^{\prime2}\left(  r\right)  }{2}\text{ and }G^{rt}=-\mathcal{L}%
^{\left(  1,0\right)  }\left(  s,0\right)  A_{t}^{\prime}\left(  r\right)
\text{.} \label{eq:SGrt}%
\end{equation}
It can show that eqn. $\left(  \ref{eq:thetathetaEOM}\right)  $ can be derived
from eqns. $\left(  \ref{eq:ttEOM}\right)  $ and $\left(  \ref{eq:NLEDEOM}%
\right)  $.

Solving eqn. $\left(  \ref{eq:NLEDEOM}\right)  $, we find that%
\begin{equation}
G^{tr}=\frac{q}{r^{2}}\text{,}\label{eq:Grt}%
\end{equation}
where $q$ is a constant. The charge of the system inside the cavity is defined
as \cite{IN-Wang:2018xdz}.%
\[
Q=-\frac{1}{4\pi}\int_{S^{2}}d^{2}x\sqrt{\sigma}n_{\mu}l_{\nu}G^{\mu\nu},
\]
where $l^{\mu}$ is the unit normal vector of the constant $t$ hypersurface,
and $\sigma$ is the induced metric on $S^{2}$. Using eqn. $\left(
\ref{eq:Grt}\right)  $, one finds that the charge inside the cavity becomes%
\begin{equation}
Q=\frac{1}{4\pi}\int d\theta d\phi r_{B}^{2}\sin\theta\frac{q}{r_{B}^{2}}=q.
\end{equation}
From eqns. $\left(  \ref{eq:SGrt}\right)  $ and $\left(  \ref{eq:Grt}\right)
$, $A_{t}^{\prime}\left(  r\right)  $ is determined by%
\begin{equation}
\mathcal{L}^{\left(  1,0\right)  }\left(  \frac{A_{t}^{\prime2}\left(
r\right)  }{2},0\right)  A_{t}^{\prime}\left(  r\right)  =\frac{Q}{r^{2}%
}.\label{eq:QAt}%
\end{equation}
The gauge potential measured on $\partial\mathcal{M}$ with respect to the
horizon is%
\begin{equation}
\Phi=4\pi\int_{r_{+}}^{r_{B}}A_{t}^{\prime}\left(  r\right)  =\frac{4\pi
A_{t}\left(  r_{B}\right)  }{\sqrt{f\left(  r_{B}\right)  }}%
,\label{eq:potential}%
\end{equation}
where the blueshift factor $1/\sqrt{f\left(  r_{B}\right)  }$ relates $A_{t}$
to the proper orthonormal frame component of the potential one-form $A$
\cite{IN-Braden:1990hw}, and we fix the gauge field $A_{t}\left(  r\right)  $
at the horizon to be zero, i.e., $A_{t}\left(  r_{+}\right)  =0$.

By integrating eqn. $\left(  \ref{eq:ttEOM}\right)  $, we have
\begin{equation}
f\left(  r\right)  =1-\frac{M}{8\pi r}-\frac{1}{2r}\int_{r}^{\infty}%
drr^{2}\left[  \mathcal{L}\left(  \frac{A_{t}^{\prime2}\left(  r\right)  }%
{2},0\right)  -A_{t}^{\prime}\left(  r\right)  \frac{Q}{r^{2}}\right]  ,
\label{eq:f(r)inf}%
\end{equation}
where $M$ is the mass of the black hole \cite{IN-Wang:2018xdz}. Suppose that
$r_{+}$ is the outer event horizon radius of the black hole. Since $f\left(
r_{+}\right)  =0$, we can express $f\left(  r\right)  $ in terms of $r_{+}$:%
\begin{equation}
f\left(  r\right)  =1-\frac{r_{+}}{r}+\frac{1}{2r}\int_{r_{+}}^{r}%
drr^{2}\mathcal{L}\left(  \frac{A_{t}^{\prime2}\left(  r\right)  }%
{2},0\right)  -\frac{Q}{2r}A_{t}\left(  r\right)  . \label{eq:f(r)}%
\end{equation}

\subsection{Euclidean Action}

In the semiclassical approximation, one can relate the on-shell Euclidean
action to the thermal partition function:%
\begin{equation}
Z\simeq e^{-\mathcal{S}^{E}},
\end{equation}
where $\mathcal{S}^{E}$ is the Euclidean continuation of the action
$\mathcal{S}$: $\mathcal{S}^{E}=i\mathcal{S}$. The Euclidean time $\tau$ is
obtained from Lorentzian time $t$ by the analytic continuation $t=i\tau$. From
$A_{\tau}d\tau=A_{t}dt$, it follows that%
\begin{equation}
A_{\tau}=-iA_{t}\text{,}%
\end{equation}
which gives $G_{r\tau}=-iG_{rt}$. So eqn. $\left(  \ref{eq:Grt}\right)  $
becomes%
\begin{equation}
G_{r\tau}=-i\frac{Q}{r^{2}}. \label{eq:Gtr}%
\end{equation}
Moreover, the gauge potential on $\partial\mathcal{M}$ is%
\begin{equation}
\Phi=\frac{4\pi iA_{\tau}\left(  r_{B}\right)  }{\sqrt{f\left(  r_{B}\right)
}}.
\end{equation}
Since the temperature $T$ is fixed on the boundary of the cavity, we can
impose the boundary condition at $r=r_{B}$ in terms of the reciprocal
temperature:%
\begin{equation}
\int\sqrt{f\left(  r_{B}\right)  }d\tau=T^{-1}.
\end{equation}
which identifies the Euclidean time $\tau$ as $\tau\sim\tau+\frac{1}%
{T\sqrt{f\left(  r_{B}\right)  }}$, and hence the period of $\tau$ is
$\frac{1}{T\sqrt{f\left(  r_{B}\right)  }}$.

For the black hole solution $\left(  \ref{eq:ansatz}\right)  $, one can
evaluate the Euclidean action by integrating over angles and performing the
integration by parts:%
\begin{align}
\mathcal{S}^{E}  &  =-\frac{8\pi}{T\sqrt{f\left(  r_{B}\right)  }}\int_{r_{+}%
}^{r_{B}}dr\left[  f\left(  r\right)  +1+rf^{\prime}\left(  r\right)  \right]
+\frac{16\pi r_{B}}{T}\nonumber\\
&  -16\pi^{2}r_{+}^{2}-\frac{4\pi}{T\sqrt{f\left(  r_{B}\right)  }}\int
_{r_{+}}^{r_{B}}drr^{2}\mathcal{L}\left(  -\frac{A_{\tau}^{\prime2}\left(
r\right)  }{2},0\right)  +\frac{4\pi iA_{\tau}\left(  r_{B}\right)  Q}%
{T\sqrt{f\left(  r_{B}\right)  }}. \label{eq:EActionf}%
\end{align}
After eqn. $\left(  \ref{eq:f(r)}\right)  $ is plugged into eqn. $\left(
\ref{eq:EActionf}\right)  $, a straightforward calculation gives%
\begin{equation}
\mathcal{S}^{E}=\frac{16\pi r_{B}}{T}\left[  1-\sqrt{f\left(  r_{B}\right)
}\right]  -S, \label{eq:EAction}%
\end{equation}
where $S=16\pi^{2}r_{+}^{2}$ is the entropy of the black hole.

For large values of $r_{B}$, one finds that%
\begin{equation}
f\left(  r_{B}\right)  =1-\frac{M}{8\pi r_{B}}+\frac{Q^{2}}{4r_{B}^{2}%
}+\mathcal{O}\left(  r_{B}^{-4}\right)  . \label{eq:f(r)LV}%
\end{equation}
In the limit of $r_{B}\rightarrow\infty$, the Euclidean action then reduces
to
\begin{equation}
\mathcal{S}^{E}=\frac{1}{T}\left(  M-TS\right)  ,
\end{equation}
as expected.

\subsection{Thermodynamics}

Various thermodynamic quantities can be derived from the Euclidean action
$\left(  \ref{eq:EAction}\right)  $, which is related to the free energy $F$
in the semiclassical approximation by%
\begin{equation}
F=-T\ln Z=T\mathcal{S}^{E}.
\end{equation}
From eqns. $\left(  \ref{eq:f(r)}\right)  $ and $\left(  \ref{eq:EAction}%
\right)  $, one finds that the free energy $F$ is a function of the
temperature $T$, the charge $Q$, the cavity radius $r_{B}$ and the horizon
radius $r_{+}$:%
\begin{equation}
F=F\left(  r_{+};T,Q,r_{B}\right)  ,\label{eq:F(rplus)}%
\end{equation}
where $T$, $Q$ and $r_{B}$ are parameters of the canonical ensemble. The only
variable $r_{+}$ can be determined by extremizing the free energy $F\left(
r_{+};T,Q,r_{B}\right)  $ with respect to $r_{+}$:%
\begin{equation}
\frac{dF\left(  r_{+};T,Q,r_{B}\right)  }{dr_{+}}=0\Longrightarrow
-\frac{d\left[  r_{B}f\left(  r_{B}\right)  \right]  /dr_{+}}{2\sqrt{f\left(
r_{B}\right)  }}=2\pi r_{+}T\Longrightarrow f^{\prime}\left(  r_{+}\right)
=4\pi T\sqrt{f\left(  r_{B}\right)  },\label{eq:frpus}%
\end{equation}
where we use $d\left[  r_{B}f\left(  r_{B}\right)  \right]  /dr_{+}%
=-r_{+}f^{\prime}\left(  r_{+}\right)  $. That said, the solution $r_{+}%
=r_{+}\left(  T,Q,r_{B}\right)  $ of eqn. $\left(  \ref{eq:frpus}\right)  $
corresponds to a locally stationary point of $F\left(  r_{+};T,Q,r_{B}\right)
$. It is interesting to note that eqn. $\left(  \ref{eq:frpus}\right)  $ can
be written as
\begin{equation}
T=\frac{T_{h}}{\sqrt{f\left(  r_{B}\right)  }}\text{,}\label{eq:TBlue}%
\end{equation}
where
\begin{equation}
T_{h}=\frac{f^{\prime}\left(  r_{+}\right)  }{4\pi}=\frac{1}{4\pi r_{+}%
}\left\{  1+\frac{r_{+}^{2}}{2}\mathcal{L}\left(  \frac{A_{t}^{\prime2}\left(
r_{+}\right)  }{2},0\right)  -\frac{A_{t}^{\prime}\left(  r_{+}\right)  Q}%
{2}\right\}  ,\label{eq:HT}%
\end{equation}
is the Hawking temperature of the black hole. The temperature $T$ on
$\partial\mathcal{M}$ is thus blueshifted from $T_{h}$, which is measured at infinity.

After obtained $r_{+}=r_{+}\left(  T,Q,r_{B}\right)  $, we can evaluate
$F\left(  r_{+};T,Q,r_{B}\right)  $ at the locally stationary point
$r_{+}=r_{+}\left(  T,Q,r_{B}\right)  $:%
\begin{equation}
F\left(  T,Q,r_{B}\right)  \equiv F\left(  r_{+}\left(  T,Q,r_{B}\right)
;T,Q,r_{B}\right)  \text{.}%
\end{equation}
For later convenience, we shall suppress $T,Q$ and $r_{B}$ in $F\left(
r_{+};T,Q,r_{B}\right)  $ and $F\left(  T,Q,r_{B}\right)  $ and denote
$F\left(  r_{+};T,Q,r_{B}\right)  $ and $F\left(  T,Q,r_{B}\right)  $ as
$F\left(  r_{+}\right)  $ and $F$, respectively. The thermal energy of the
black hole in the cavity is%
\begin{equation}
E=-T^{2}\frac{\partial\left(  F/T\right)  }{\partial T}=16\pi r_{B}\left[
1-\sqrt{f\left(  r_{B}\right)  }\right]  .
\end{equation}
Using eqn. $\left(  \ref{eq:f(r)inf}\right)  $, we can express the ADM mass of
the black hole $M$ in terms of $E$ and $Q$:%
\begin{equation}
M=E-\frac{E^{2}}{32\pi r_{B}}-4\pi\int_{r_{B}}^{\infty}drr^{2}\left[
\mathcal{L}\left(  \frac{A_{t}^{\prime2}\left(  r\right)  }{2},0\right)
-A_{t}^{\prime}\left(  r\right)  \frac{Q}{r^{2}}\right]  ,
\end{equation}
where the second and third terms on left-hand side can be interpreted as the
gravitational and electrostatic binding energies, respectively. Using eqn.
$\left(  \ref{eq:f(r)}\right)  $, we can express the thermal energy $E$ in
terms of the entropy $S$, the charge $Q$ and the cavity radius $r_{B}$.
Differentiating $E$ with respect to $S$ and $Q$, respectively, gives
\begin{align}
\frac{\partial E}{\partial S} &  =-\frac{d\left[  r_{B}f\left(  r_{B}\right)
\right]  /dr_{+}}{4\pi r_{+}\sqrt{f\left(  r_{B}\right)  }}=T,\nonumber\\
\frac{\partial E}{\partial Q} &  =-\frac{8\pi r_{B}}{\sqrt{f\left(
r_{B}\right)  }}\frac{\partial f\left(  r_{B}\right)  }{\partial Q}%
=\Phi.\label{eq:ESQ}%
\end{align}
From the energy $E$, we can define a thermodynamic surface pressure by%
\begin{equation}
\lambda\equiv-\frac{\partial E}{\partial\left(  4\pi r_{B}^{2}\right)  }%
=\frac{\left(  1-\sqrt{f\left(  r_{B}\right)  }\right)  ^{2}}{r_{B}%
\sqrt{f\left(  r_{B}\right)  }}+\frac{r_{B}^{2}\mathcal{L}\left(  \frac
{A_{t}^{\prime2}\left(  r_{B}\right)  }{2},0\right)  -QA_{t}^{\prime}\left(
r_{B}\right)  }{2r_{B}\sqrt{f\left(  r_{B}\right)  }}.\label{eq:lamda}%
\end{equation}
From eqns. $\left(  \ref{eq:ESQ}\right)  $ and $\left(  \ref{eq:lamda}\right)
$, the first law of thermodynamics can be established:%
\begin{equation}
dE=TdS+\Phi dQ-\lambda dA,
\end{equation}
where $A\equiv4\pi r_{B}^{2}$ is the surface area of the cavity.

To obtain the proper Smarr relation for the black hole, we need to consider
the dimensionful couplings $a_{i}$ in the NLED Lagrangian $\mathcal{L}\left(
s,p\right)  $, which have $\left[  a_{i}\right]  =L^{c_{i}}$. Since the
dimensional analysis give that $\left[  Q\right]  =\left[  M\right]  =L$,
$\left[  T\right]  =\left[  \lambda\right]  =L^{-1}$and $\left[  A\right]
=\left[  S\right]  =L^{2}$, the Euler scaling argument leads to the Smarr
relation%
\begin{equation}
M=2\left(  TS-\lambda A\right)  +\sum\limits_{i}c_{i}a_{i}\mathcal{A}%
_{i}+Q\Phi,
\end{equation}
where we introduce the conjugates $\mathcal{A}_{i}$ associated with $a_{i}$:%
\begin{equation}
\mathcal{A}_{i}=\frac{\partial M}{\partial a_{i}}.
\end{equation}

We now discuss the thermodynamic stability of the black hole in the cavity
against thermal fluctuations. In the canonical ensemble, one considers the
specific heat at constant electric charge:%
\begin{equation}
C_{Q}=T\left(  \frac{\partial S}{\partial T}\right)  _{Q}=32\pi^{2}%
r_{+}\left(  T,Q,r_{B}\right)  T\frac{\partial r_{+}\left(  T,Q,r_{B}\right)
}{\partial T}.
\end{equation}
When $C_{Q}>0$, the system is thermally stable. Thus, a thermally stable black
hole phase has $\frac{\partial r_{+}\left(  T,Q,r_{B}\right)  }{\partial T}%
>0$. Since $\partial^{2}F/\partial^{2}T=-C_{Q}$, the thermally stable/unstable
phases have concave downward/upward $F$-$T$ curves. On the other hand, it can
show that, at $r_{+}=r_{+}\left(  T,Q,r_{B}\right)  $,
\begin{equation}
\frac{\partial^{2}F\left(  r_{+}\right)  }{\partial r_{+}^{2}}=\frac{32\pi
^{2}r_{+}}{\partial r_{+}/\partial T}\text{,}%
\end{equation}
which means that the black hole phase is thermally stable/unstable if
$r_{+}\left(  T,Q,r_{B}\right)  $ is a local minimum/maximum of $F\left(
r_{+}\right)  $.

To find the global minimum of $F\left(  r_{+}\right)  $ over the space of the
variable $r_{+}$ with fixed values of $T,Q$ and $r_{B}$, we also need to
consider the values of $F\left(  r_{+}\right)  $ at the edges of the space of
$r_{+}$. In fact, the physical space of $r_{+}$ is constrained by%
\begin{equation}
r_{e}\leq r_{+}\leq r_{B}\text{,}%
\end{equation}
where $r_{e}$ is the horizon radius of the extremal black hole with the charge
being $Q$. If there exists no extremal black hole solution for $Q$, one can
simply set $r_{e}=0$. For simplicity, the global minimum of $F\left(
r_{+}\right)  $ at the edges is dubbed "edge state (ES)" in our paper.

\section{Born-Infeld Black Hole in a Cavity}

\label{Sec:BIBHC}

BI electrodynamics is described by the Lagrangian density%
\begin{equation}
\mathcal{L}\left(  s,p\right)  =\frac{1}{a}\left(  1-\sqrt{1-2as}\right)
\text{,} \label{eq:BI}%
\end{equation}
where the coupling parameter $a$ is related to the string tension
$\alpha^{\prime}$ as $a=\left(  2\pi\alpha^{\prime}\right)  ^{2}>0$. For
$a=0$, the BI Lagrangian would reduce to the Maxwell Lagrangian. Solving eqn.
$\left(  \ref{eq:QAt}\right)  $ for $A_{t}^{\prime}\left(  r\right)  $ gives
\begin{equation}
A_{t}^{\prime}\left(  r\right)  =\frac{Q}{\sqrt{r^{4}+aQ^{2}}},
\end{equation}
where $Q$ is the charge of the BI black hole. From eqn. $\left(
\ref{eq:f(r)}\right)  $, one can express $f\left(  r\right)  $ in terms of the
horizon radius $r_{+}$:
\begin{align}
f\left(  r\right)   &  =1-\frac{r_{+}}{r}+\frac{r_{+}}{r}\frac{Q^{2}}%
{6\sqrt{r_{+}^{4}+aQ^{2}}+6r_{+}^{2}}-\frac{Q^{2}}{3rr_{+}}\text{ }_{2}%
F_{1}\left(  \frac{1}{4},\frac{1}{2},\frac{5}{4};-\frac{aQ^{2}}{r_{+}^{4}%
}\right) \nonumber\\
&  -\frac{Q^{2}}{6\sqrt{r^{4}+aQ^{2}}+6r^{2}}+\frac{Q^{2}}{3r^{2}}\text{ }%
_{2}F_{1}\left(  \frac{1}{4},\frac{1}{2},\frac{5}{4};-\frac{aQ^{2}}{r^{4}%
}\right)  , \label{eq:BIBHf(r)}%
\end{align}
where $_{2}F_{1}\left(  a,b,c;x\right)  $ is the hypergeometric function.

It is convenient to express quantities in units of $r_{B}$:%
\begin{equation}
x\equiv\frac{r_{+}}{r_{B}}\text{, }\tilde{Q}\equiv\frac{Q}{r_{B}}\text{,
}\tilde{a}\equiv\frac{a}{r_{B}^{2}}\text{, }\tilde{T}\equiv r_{B}T\text{,
}\tilde{F}\left(  x\right)  =\frac{F\left(  r_{+}\right)  }{16\pi r_{B}}\text{
and }\tilde{F}=\frac{F}{16\pi r_{B}}\text{,}%
\end{equation}
where $r_{+}$ is the horizon radius. We then use eqns. $\left(
\ref{eq:EAction}\right)  $ and $\left(  \ref{eq:BIBHf(r)}\right)  $ to find
the free energy as a function of $x$:%
\begin{equation}
\tilde{F}\left(  x\right)  =1-\sqrt{f\left(  x\right)  }-\pi x^{2}\tilde{T},
\end{equation}
where
\begin{align}
f\left(  x\right)   &  =1-x+\frac{x\tilde{Q}^{2}}{6\sqrt{x^{4}+\tilde{a}%
\tilde{Q}^{2}}+6x^{2}}-\frac{\tilde{Q}^{2}}{3x}\text{ }_{2}F_{1}\left(
\frac{1}{4},\frac{1}{2},\frac{5}{4};-\frac{\tilde{a}\tilde{Q}^{2}}{x^{4}%
}\right) \nonumber\\
&  -\frac{\tilde{Q}^{2}}{6\sqrt{1+\tilde{a}\tilde{Q}^{2}}+6}+\frac{\tilde
{Q}^{2}}{3}\text{ }_{2}F_{1}\left(  \frac{1}{4},\frac{1}{2},\frac{5}%
{4};-\tilde{a}\tilde{Q}^{2}\right)  .
\end{align}
The Hawking temperature of the BI black hole can be calculated from eqn.
$\left(  \ref{eq:HT}\right)  $:%

\begin{equation}
\tilde{T}_{h}\equiv r_{B}T_{h}=\frac{1}{4\pi x}\left(  1-\frac{1}{2}%
\frac{\tilde{Q}^{2}}{x^{2}+\sqrt{x^{4}+\tilde{a}\tilde{Q}^{2}}}\right)  .
\label{eq:HTBIBH}%
\end{equation}
The locally stationary points of $\tilde{F}\left(  x\right)  $ are determined
by $d\tilde{F}\left(  x\right)  /dx=0$, which becomes%
\begin{equation}
\tilde{T}=\frac{\tilde{T}_{h}}{\sqrt{f\left(  x\right)  }}. \label{eq:TBIBH}%
\end{equation}

As shown in \cite{IN-Tao:2017fsy}, there are two types of BI black holes
depending on the minimum value of $\tilde{T}_{h}$:

\begin{itemize}
\item RN type: $4\tilde{a}\leq\tilde{Q}^{2}\leq4\left(  1+\tilde{a}\right)  $.
This type of BI black holes can have extremal black hole solutions like RN
black holes. In fact, the Hawking temperature $\tilde{T}_{h}=0$ has one single
solution $x=x_{e}\equiv\frac{1}{2}\sqrt{\tilde{Q}^{2}-4\tilde{a}}$, where
$x_{e}r_{B}$ is the horizon radius of the extremal BI black hole with $Q$ and
$a$. In this case, we must have $x_{e}\leq x\leq1$. Note that requiring
$x_{e}\leq1$ puts an upper bound on $\tilde{Q}^{2}$: $\tilde{Q}^{2}%
\leq4\left(  1+\tilde{a}\right)  $. Another way to understand this upper bound
is that, when $\tilde{Q}^{2}>4\left(  1+\tilde{a}\right)  $, $f\left(
x\right)  $ is always negative and hence $\tilde{F}\left(  x\right)  $ is not
a real-valued function for any $x$. When $x$ $=x_{e}$ we have an extremal BI
black hole, and when $x=1$ the horizon merges with the boundary.

\item Schwarzschild-like type: $\tilde{Q}^{2}<4\tilde{a}$. This type of BI
black holes have only one horizon like Schwarzschild black holes. The Hawking
temperature $\tilde{T}_{h}$ has a positive minimum value and goes to $+\infty$
as $x\rightarrow0$. In this case, we can have $0\leq x\leq1$, over which
$\tilde{F}\left(  x\right)  $ is well-defined. It can show that $\tilde{T}%
_{h}/\sqrt{f\left(  x\right)  }$ has a minimum value of $\tilde{T}_{\min}>0$
over $0\leq x\leq1$. Eqn. $\left(  \ref{eq:TBIBH}\right)  $ implies that the
locally stationary points of $\tilde{F}\left(  x\right)  $, corresponding to
BI black hole solutions, only exist for $\tilde{T}\geq\tilde{T}_{\min}$. When
$x$ $=0$, one finds that eqn. $\left(  \ref{eq:BIBHf(r)}\right)  $ becomes%
\begin{equation}
f\left(  r\right)  =1-\frac{Q^{3/2}\Gamma\left(  1/4\right)  \Gamma\left(
5/4\right)  }{3a^{1/4}\sqrt{\pi}r}-\frac{Q^{2}}{6\sqrt{r^{4}+aQ^{2}}+6r^{2}%
}+\frac{Q^{2}}{3r^{2}}\text{ }_{2}F_{1}\left(  \frac{1}{4},\frac{1}{2}%
,\frac{5}{4};-\frac{aQ^{2}}{r^{4}}\right)  .
\end{equation}
If $Q=0$, $f\left(  r\right)  =1$, and hence the edge state at $x=0$ is just
the thermal flat space. For $Q>0$, we have%
\begin{equation}
f\left(  r\right)  =1-\frac{Q}{2\sqrt{a}}+\mathcal{O}\left(  r\right)  \text{
and }R=\frac{Q}{\sqrt{a}r^{2}}-\frac{2}{a}+\mathcal{O}\left(  r\right)
\text{,}%
\end{equation}
where $R$ is the Ricci scalar. So the metric has a physical singularity at
$r=0$ although $f\left(  0\right)  $ is finite. It can show that $f\left(
r\right)  >0$, and hence there exists no horizon. The edge state with $Q>0$ at
$x=0$ is thus a naked singularity.
\end{itemize}

\begin{figure}[tb]
\begin{center}
\includegraphics[width=0.48\textwidth]{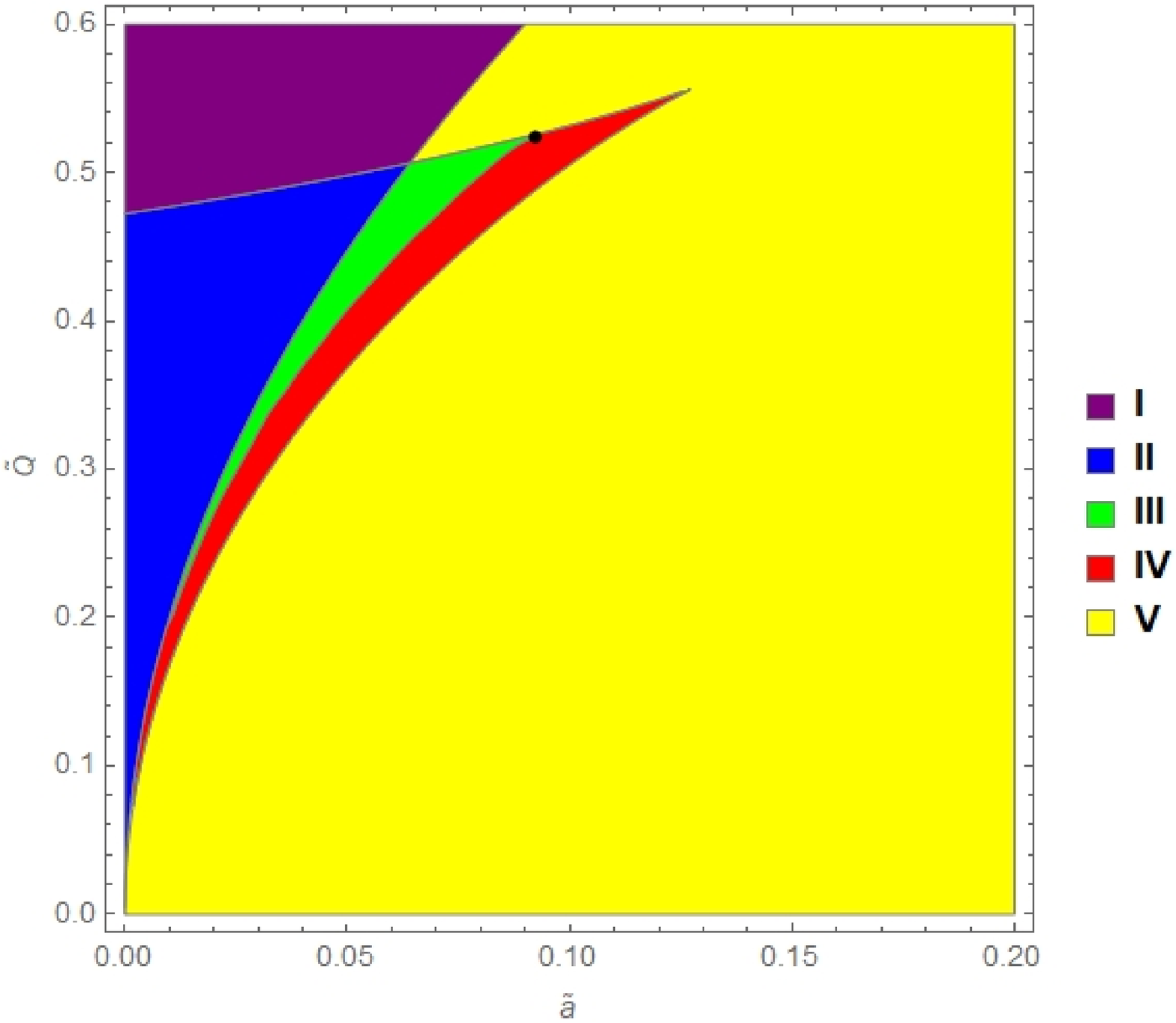}
\includegraphics[width=0.46\textwidth]{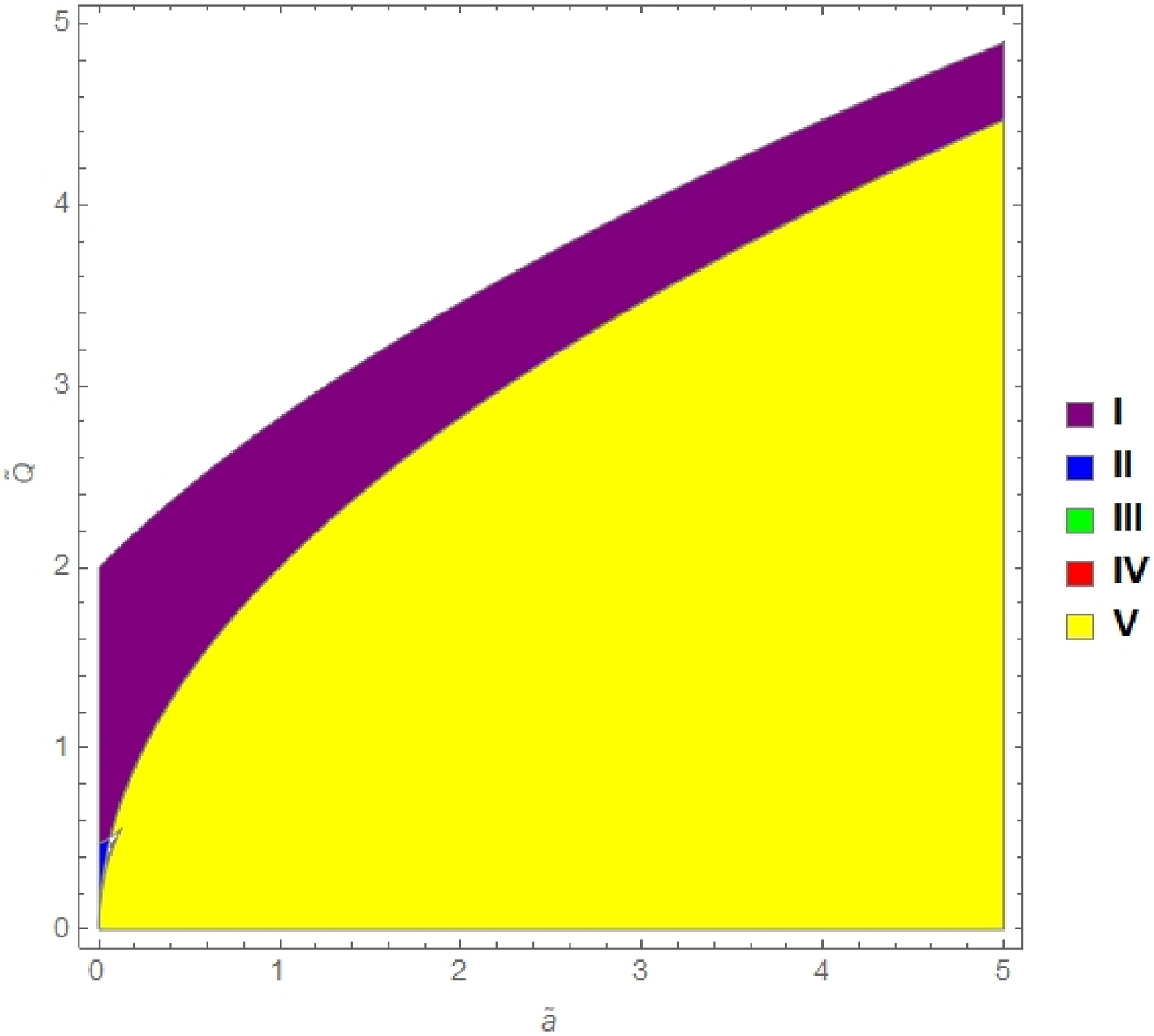}
\end{center}
\caption{{\footnotesize The five regions in the $\tilde{a}$-$\tilde{Q}$ phase
space of a BI black hole (BH) in a cavity, each of which possesses distinct
behavior of the phase structure and transition. Varying the temperature, a van
der Waals-like LBH/SBH phase transition occurs in Regions II and III while a
Hawking-Page-like ES/BH phase transition occurs in Regions III, IV and V.
There is only one phase in Regions I. The critical line consists of $\tilde
{Q}_{12}\left(  \tilde{a}\right)  $ and $\tilde{Q}_{35}\left(  \tilde
{a}\right)  $, where $\tilde{Q}_{ij}\left(  \tilde{a}\right)  $ is the
boundary Region $i$ and Region $j$. The critical line terminates at the black
dot, which is at $\left\{  \tilde{a}_{c},\tilde{Q}_{c}\right\}  \approx
\left\{  0.092,0.524\right\}  $.}}%
\label{fig:RP}%
\end{figure}

\begin{figure}[tb]
\begin{center}
\includegraphics[width=0.5\textwidth]{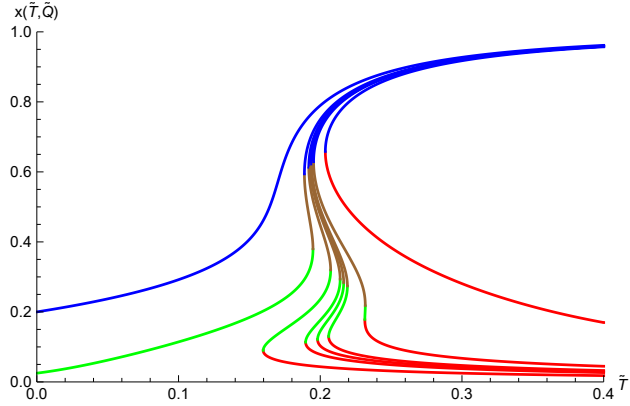}
\end{center}
\caption{{\footnotesize Plots of $x(\tilde{T},\tilde{Q})$ against $\tilde{T}$
for various values of $\tilde{Q}$ with $\tilde{a}=0.05$, where $x(\tilde
{T},\tilde{Q})$ is the locally stationary point of $\tilde{F}\left(  x\right)
$. From left to right, $\tilde{Q}=0.600$ (Region I), $0.450$ (Region II),
$0.415$ (Region III), $0.400$ (Region IV), $0.395$ (Region IV), $0.390$
(Region IV), $0.370$ (Region IV) and $0.200$ (Region V). Since thermally
stable phases have $\partial x/\partial\tilde{T}>0$, the BI BHs on blue and
green segments of the curves are thermally stable. We denote the BI BHs on
blue segments by Large BH (or BH if there is no green segment on the curve)
and these on green segments by Small BH. The BI BHs on red and brown segments
are denoted by Intermediate BH, which are thermally unstable.}}%
\label{fig:xvsT}%
\end{figure}\begin{figure}[tbh]
\begin{center}
\subfigure[{~\scriptsize Region I: $\tilde{a}=0.01$ and $\tilde{Q}=0.6$. There is no phase transition.}]{
\includegraphics[width=0.9\textwidth]{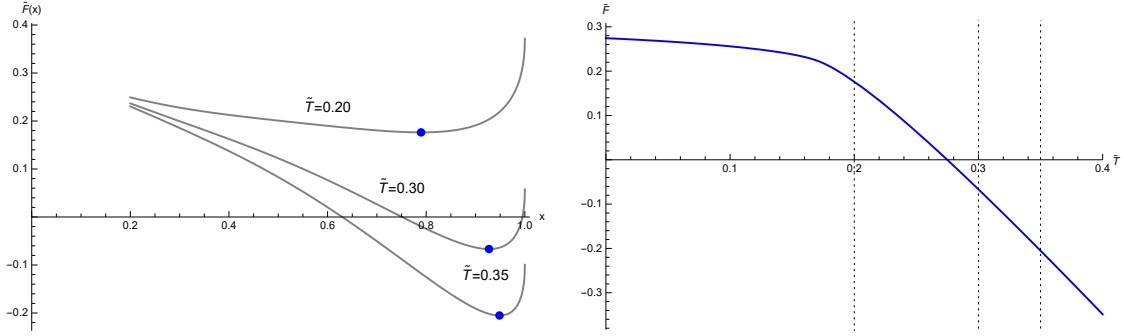}\label{fig:R12:a}}
\subfigure[{~\scriptsize Region II: $\tilde{a}=0.01$ and $\tilde{Q}=0.3$. There is a first-order phase
transition between Small BH and Large BH.}]{
\includegraphics[width=0.9\textwidth]{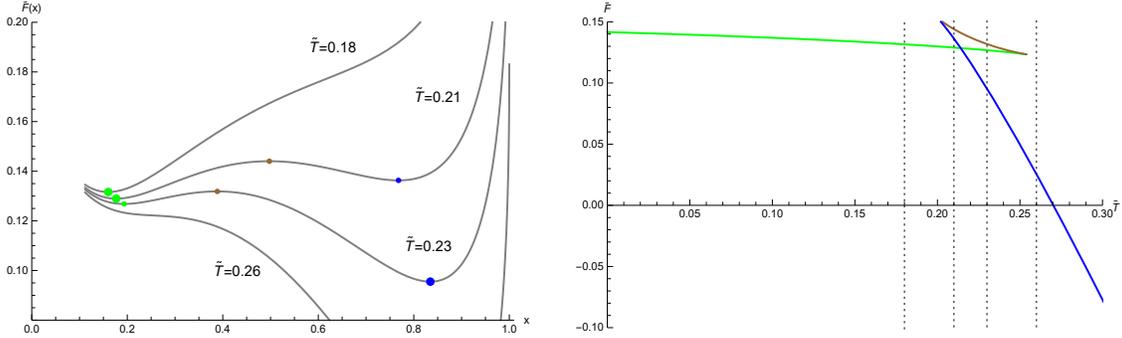}\label{fig:R12:b}}
\end{center}
\caption{{\footnotesize Plots of the free energies $\tilde{F}\left(  x\right)
$ and $\tilde{F}$ against $\tilde{T}$ for the BI BHs in Regions I and II. The
BI BHs in these regions are RN type. The BI BHs on the blue and green branches
(dots) are thermally stable. \textbf{Left Panels}: $\tilde{F}\left(  x\right)
$ is plotted for various values of $\tilde{T}$, which are depicted by the
vertical black dotted lines in the right panels. The locally stationary points
of $\tilde{F}\left(  x\right)  $ are marked with colored dots, the colors of
which correspond to these of segments in FIG. \ref{fig:xvsT}. Larger dots
represent global minimums of $\tilde{F}\left(  x\right)  $, which are globally
stable. \textbf{Right Panels}: The values of $\tilde{F}\left(  x\right)  $
evaluated at the locally stationary points are plotted against $\tilde{T}$.
Their colors match these in the left panels and FIG. \ref{fig:xvsT}, which
means that blue/green branches are Large/Small BH, and the brown branch is
Intermediate BH.}}%
\end{figure}

\begin{figure}[tb]
\begin{center}
\includegraphics[width=0.9\textwidth]{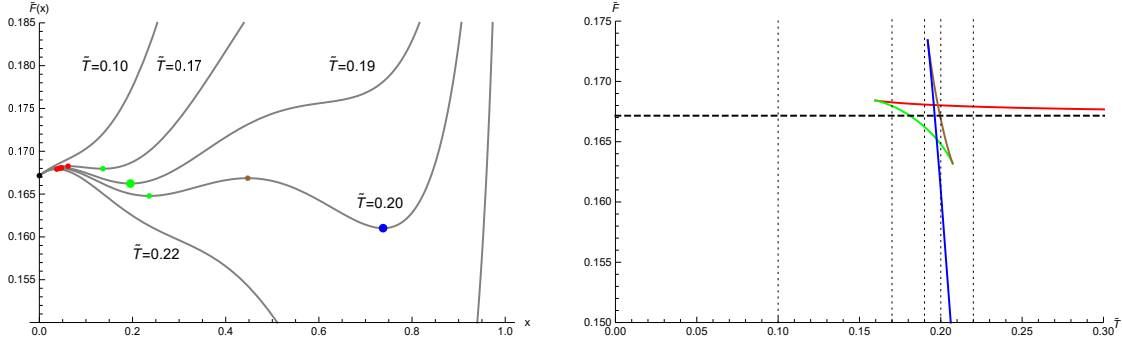}
\end{center}
\caption{{\footnotesize Region III: $\tilde{a}=0.05$ and $\tilde{Q}=0.415$.
The BI BHs in this region are Schwarzschild-like type, which implies that $x$
can go to $0$. \textbf{Left Panel}: The locally stationary points of
$\tilde{F}\left(  x\right)  $ are represented with colored dots, and
$\tilde{F}\left(  0\right)  $ is marked with the black dot. Larger colored
dots represent global minimums of $\tilde{F}\left(  x\right)  $. If there is
no larger colored dot for some $\tilde{T}$, $\tilde{F}\left(  x\right)  $ has
the global minimum at $x=0$. \textbf{Right Panel}: The colored branches are
the values of $\tilde{F}\left(  x\right)  $ evaluated at the locally
stationary points as functions of $\tilde{T}$. Specifically, the blue/green
branches are Large/Small BH, and the brown/red branches are Intermediate BH.
The horizontal black dashed line is $\tilde{F}\left(  0\right)  $. As
$\tilde{T}$ increases from zero, a first-order phase transition from the edge
state to Small BH occurs at some $\tilde{T}$. Further increasing $\tilde{T}$,
there would be another first-order phase transition from Small BH to Large
BH.}}%
\label{fig:R3}%
\end{figure}

\begin{figure}[ptb]
\begin{center}
\subfigure[{~ \scriptsize Region IV: $\tilde{a}=0.05$. Top Panels: $\tilde{Q}=0.400$. Bottom-Left Panel:
$\tilde{Q}=0.395$. Bottom-Right Panel: $\tilde{Q}=0.390$.}]{
\includegraphics[width=0.9\textwidth]{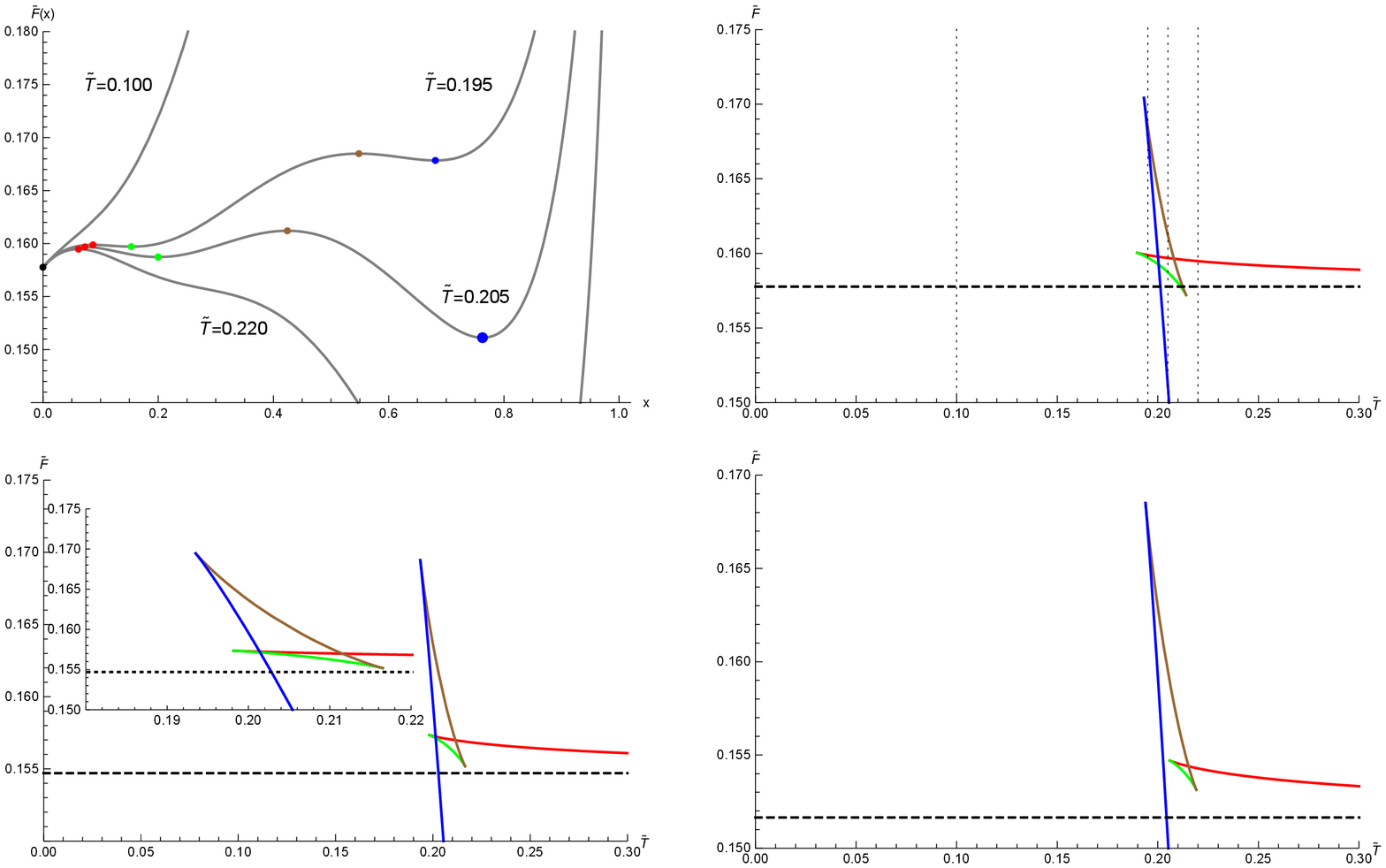}\label{fig:R45:a}}
\subfigure[{~ \scriptsize Region V: $\tilde{a}=0.05$ and $\tilde{Q}=0.4$.}]{
\includegraphics[width=0.9\textwidth]{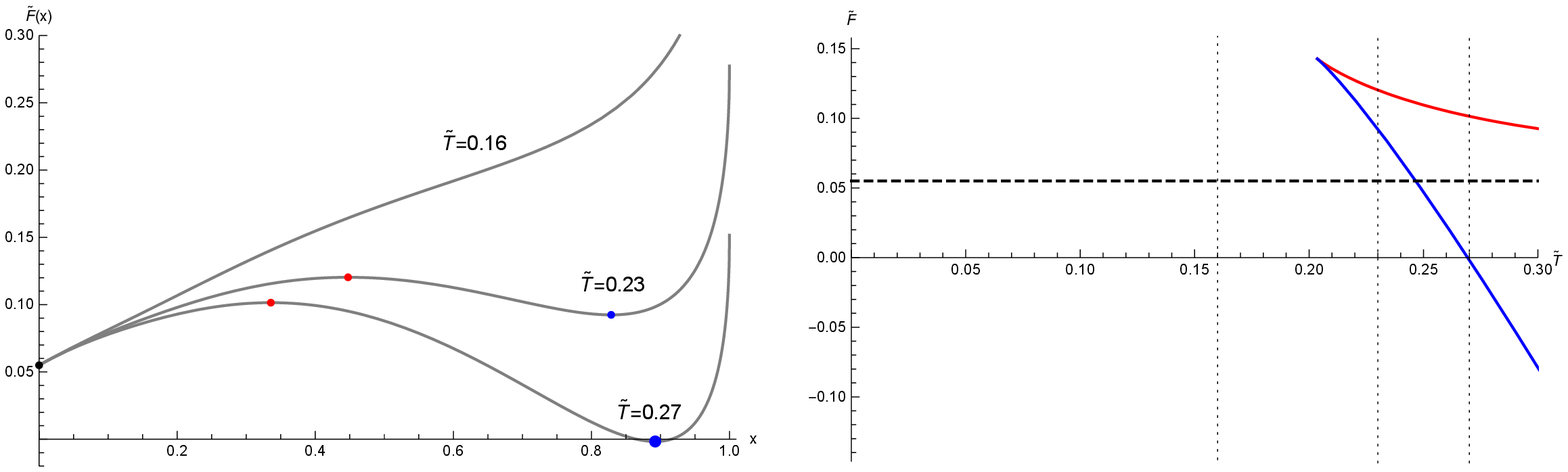}\label{fig:R45:b}}
\end{center}
\caption{{\footnotesize Plots of the free energies $\tilde{F}\left(  x\right)
$ and $\tilde{F}$ against $\tilde{T}$ for the BI BHs in Regions IV and V,
which are Schwarzschild-like type. The blue/green branches are Large/Small BH,
and the brown/red branches are Intermediate BH. The horizontal black dashed
line is $\tilde{F}\left(  0\right)  $. As $\tilde{T}$ increases from zero,
there is always a single first-order phase transition from the edge state to
Large BH occurring at some temperature.}}%
\label{fig:R45}%
\end{figure}

To find the phase structure and transition of a BI black hole in a cavity, we
need to analyze the locally stationary points of $\tilde{F}\left(  x\right)  $
and find the global minimum value of $\tilde{F}\left(  x\right)  $. In fact,
with fixed value of $\tilde{a}$, the locally stationary points of $\tilde
{F}\left(  x\right)  $ can be determined by solving eqns. $\left(
\ref{eq:TBlue}\right)  $ and $\left(  \ref{eq:HTBIBH}\right)  $ for $x$ in
terms of $\tilde{T}$ and $\tilde{Q}$. The solution $x(\tilde{T},\tilde{Q})$ is
often a multivalued function, each branch of which corresponds to a family of
BI black hole solutions. When $x(\tilde{T},\tilde{Q})$ is multivalued, there
is more than one family of BI black holes of different sizes with fixed values
of $\tilde{T}$ and $\tilde{Q}$. We find that there are five regions in the
$\tilde{a}$-$\tilde{Q}$ phase space of the BI black hole, in each of which the
BI black hole has distinct behavior of the branches of $x(\tilde{T},\tilde
{Q})$ and phase structure. The five regions of the $\tilde{a}$-$\tilde{Q}$
phase space are mapped in FIG. \ref{fig:RP}. We plot $x(\tilde{T},\tilde{Q})$
against $\tilde{T}$ for various values of $\tilde{Q}$ with $\tilde{a}=0.05$ in
FIG. \ref{fig:xvsT}, which shows the general behavior of $x(\tilde{T}%
,\tilde{Q})$ in these five regions. In what follows, we discuss the phase
structure and transition of the BI black hole in the five regions:

\begin{itemize}
\item Region I: As shown in FIG. \ref{fig:xvsT}, there is only one branch for
$x(\tilde{T},\tilde{Q})$ with fixed values of $\tilde{a}$ and $\tilde{Q}$, on
which the BI black hole is thermally stable. Since BI black holes in this
region satisfy $\tilde{Q}\geq4\tilde{a}$, they are RN type, and hence
$\tilde{T}$ can go to zero. For the BI black hole with $\tilde{a}=0.01$ and
$\tilde{Q}=0.6$ in this region, we plot the free energy $\tilde{F}\left(
x\right)  $ in FIG. \ref{fig:R12:a}, which shows that the endpoints always
have higher free energy, and the local minimum of $\tilde{F}\left(  x\right)
$ is also the global minimum. There is only one phase in this region.

\item Region II: BI black holes in this region are RN type. As shown in FIG.
\ref{fig:xvsT}, there are three branches for $x(\tilde{T},\tilde{Q})$ with
fixed values of $\tilde{a}$ and $\tilde{Q}$, which are denoted by Small BH
(green), Large BH (blue) and Intermediate\ BH (brown). Both the Small BH and
Large BH branches are thermally stable. For the BI black hole with $\tilde
{a}=0.01$ and $\tilde{Q}=0.3$ in this region, we plot the free energy
$\tilde{F}\left(  x\right)  $ in FIG. \ref{fig:R12:b}, which shows that the
endpoints always have higher free energy\ than the global minimum. $\tilde
{F}\left(  x\right)  $ has the global minimum at Small BH for small enough
$\tilde{T}$ and Large BH for large enough $\tilde{T}$, respectively. The free
energies of the three branches are plotted in the right panel of FIG.
\ref{fig:R12:b}, which shows that there is a first-order phase transition
between Small BH and Large BH.

\item Region III: BI black holes in this region are Schwarzschild-like type.
For $\tilde{T}<\tilde{T}_{\min}$, $\tilde{F}\left(  x\right)  $ is a strictly
increasing function (see $\tilde{T}=0.10$ in the left panel of FIG.
\ref{fig:R3}), and hence $\tilde{F}\left(  x\right)  $ has the global minimum
at $x=0$, dubbed the edge state. For $\tilde{T}\geq\tilde{T}_{\min}$, as shown
in FIG. \ref{fig:xvsT}, there are four branches for $x(\tilde{T},\tilde{Q})$
with fixed values of $\tilde{a}$ and $\tilde{Q}$, which are denoted by Small
BH (green), Large BH (blue) and Intermediate\ BH (brown and red). Both the
Small BH and Large BH branches are thermally stable. The free energies of the
four branches and the edge state are plotted in the right panel of FIG.
\ref{fig:R3}. As $\tilde{T}$ increases from $\tilde{T}_{\min}$, the free
energy of Small BH decrease while $\tilde{F}\left(  0\right)  $ is constant.
They cross each other at some point, where a first-order transition occurs,
and Small BH becomes globally stable. Further increasing $\tilde{T}$, Large BH
appears, and its free energy decrease more rapidly than that of Small BH. So
they cross each other at some point, where another first-order transition
occurs, and Large BH then becomes the globally stable one.

\item Region IV: BI black holes in this region are also Schwarzschild-like
type. As in Region III, the edge state at $x=0$ is globally stable for
$\tilde{T}<\tilde{T}_{\min}$. For $\tilde{T}\geq\tilde{T}_{\min}$, there are
also four branches for $x(\tilde{T},\tilde{Q})$ with fixed values of
$\tilde{a}$ and $\tilde{Q}$, which are denoted by Small BH (green), Large BH
(blue) and Intermediate\ BH (brown and red). The free energies of the four
branches and the edge state are plotted in FIG. \ref{fig:R45:a}. Unlike Region
III, the temperature at which Small BH appears is too high, such that
$\tilde{F}\left(  0\right)  $ does not cross the free energy of Small BH or
$\tilde{F}\left(  0\right)  $ crosses the free energy of Large BH\ before it
crosses that of Small BH. Hence as $\tilde{T}$ increases, there is only one
first-order transition from the edge state to Large BH at some temperature,
where the free energy of Large BH and $\tilde{F}\left(  0\right)  $ cross each other.

\item Region V: BI black holes in this region are also Schwarzschild-like
type. As in Regions III and IV, the edge state at $x=0$ is globally stable for
$\tilde{T}<\tilde{T}_{\min}$. However for $\tilde{T}\geq\tilde{T}_{\min}$,
there are two branches for $x(\tilde{T},\tilde{Q})$ with fixed values of
$\tilde{a}$ and $\tilde{Q}$, which are denoted by BH (green) and
Intermediate\ BH (red). The BH branch is thermally stable. The free energies
of the two branches and the edge state are plotted in FIG. \ref{fig:R45:b},
which shows that there is a first-order transition from the edge state to
Large BH as $\tilde{T}$ increases.
\end{itemize}

In FIG. \ref{fig:RP}, the boundary between the region in which $x(\tilde
{T},\tilde{Q})$ has $n$ branches and that in which $x(\tilde{T},\tilde{Q})$
has $n+2$ branches is the critical line. There are 3 such boundaries in FIG.
\ref{fig:RP}, i.e., $\tilde{Q}_{12}\left(  \tilde{a}\right)  $, $\tilde
{Q}_{35}\left(  \tilde{a}\right)  $ and $\tilde{Q}_{45}\left(  \tilde
{a}\right)  $, where $\tilde{Q}_{ij}\left(  \tilde{a}\right)  $ is the
boundary between Region $i$ and Region $j$. However, FIG. \ref{fig:R45} shows
that $\tilde{Q}_{45}\left(  \tilde{a}\right)  $ is not physical since it does
not globally minimize the free energy. Thus physical critical line only
consists of $\tilde{Q}_{12}\left(  \tilde{a}\right)  $ and $\tilde{Q}%
_{35}\left(  \tilde{a}\right)  $, which terminates at $\left\{  \tilde{a}%
_{c},\tilde{Q}_{c}\right\}  \simeq\left\{  0.092,0.524\right\}  $. The line
$\tilde{Q}_{12}\left(  \tilde{a}\right)  $ is reminiscent of RN black holes.

\begin{figure}[tb]
\begin{center}
\includegraphics[width=0.48\textwidth]{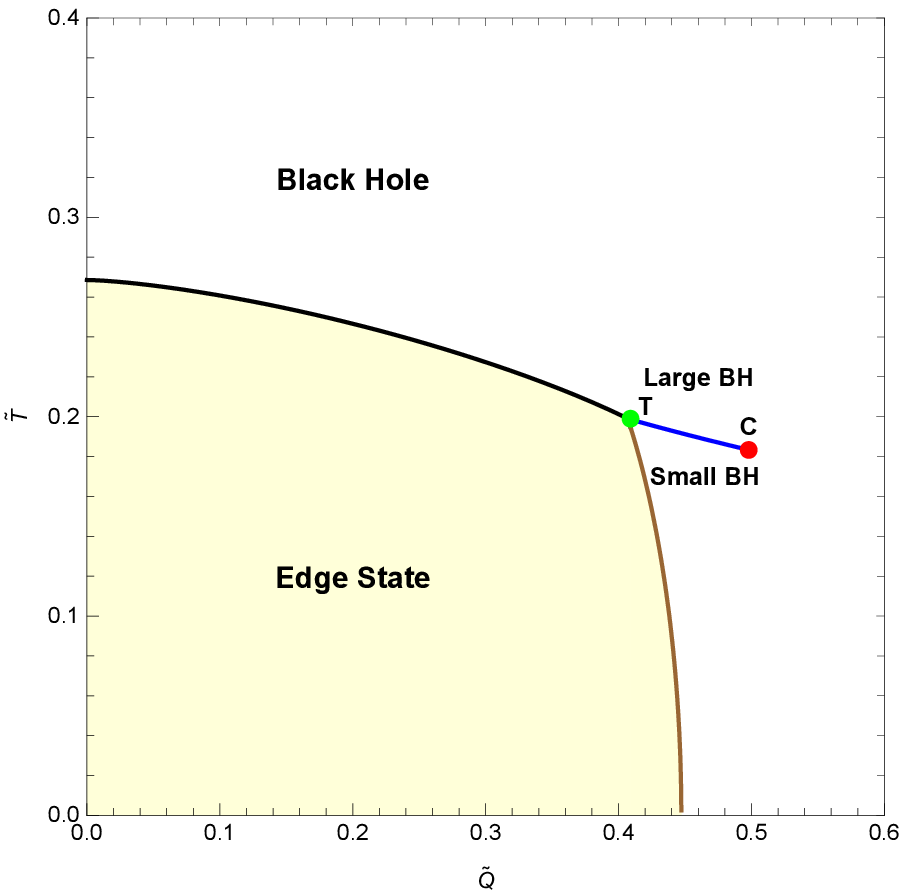}
\includegraphics[width=0.48\textwidth]{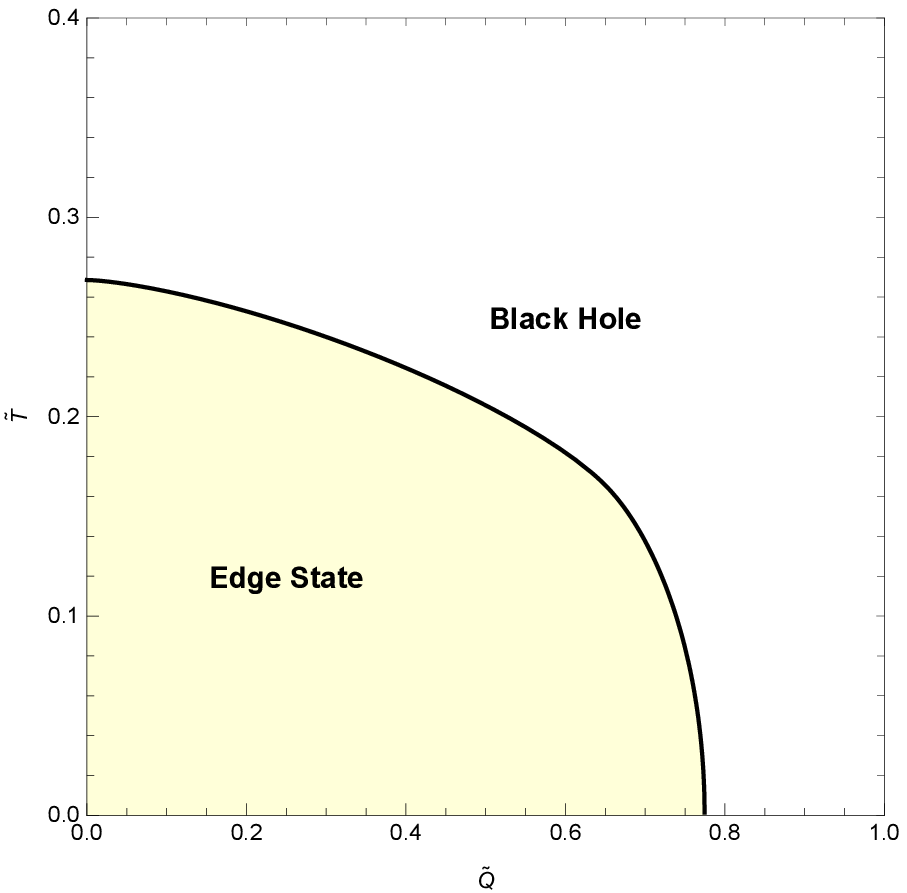}
\end{center}
\caption{{\footnotesize The phase diagrams in the $\tilde{Q}$-$\tilde{T}$
phase space. \textbf{Left Panel}: $\tilde{a}=0.05<\tilde{a}_{c}$. The
first-order phase transition line separating Large BH and Small BH is
displayed by the blue line, and it terminates at the critical point, marked by
the red dot and annotated by C. There are also a first-order phase transition
line between BH and the edge state, depicted by the black line, and a
first-order phase transition line between SBH and the edge state, depicted by
the brown line. These first-order phase transition lines merge together at the
tricritical point, marked by the green dot and annotated by T. \textbf{Right
Panel}: $\tilde{a}=0.15>\tilde{a}_{c}$. There is a first-order phase
transition line separating BH and the edge state, displayed by the black
line.}}%
\label{fig:PT}%
\end{figure}

The phase diagram with $\tilde{a}=0.05$ in the $\tilde{Q}$-$\tilde{T}$ phase
space is displayed in the left panel of FIG. \ref{fig:PT}. There is a LBH/SBH
first-order phase transition for some range of $\tilde{Q}$, a
Hawking-Page-like ES/SBH first-order phase transition for some smaller range
of $\tilde{Q}$ and another Hawking-Page-like ES/BH first-order phase
transition for some larger range of $\tilde{Q}$. The LBH/SBH phase transition
line is a van der Waals-like phase transition and hence terminates at the
critical point, represented by the red dot. These three first-order phase
transition lines merge together at the tricritical point, marked by the green
dot. The phase diagram with $\tilde{a}=0.15$ in the $\tilde{Q}$-$\tilde{T}$
phase space is plotted in the right panel of FIG. \ref{fig:PT}, which is
simpler than that with $\tilde{a}=0.05$. Since $\tilde{a}=0.15>\tilde{a}_{c}$,
there is no critical point or tricritical point in this phase diagram. There
are two phases, namely the edge state and BH, which are separated by a
Hawking-Page-like first-order phase transition line.

\section{Discussion and Conclusion}

\label{Sec:CON}

In the first part of this paper, we calculated the Euclidean action of a
general NLED black hole in a finite spherical cavity and investigated the
corresponding thermodynamic behavior in a canonical ensemble. Specifically,
the Euclidean action was given by eqn. $\left(  \ref{eq:EAction}\right)  $,
which could be interpreted as the free energy of the black hole. It was then
demonstrated that the first law of thermodynamics and the Smarr relation were
satisfied at the locally stationary points of the free energy. It also showed
that the local minimum of the free energy corresponds to the locally stable
phase of the system. To determine the globally stable phase, the edge states
are needed to be considered as well.

In the second part, we examined the phase structure and transition of a BI
black hole in a cavity. In FIG. \ref{fig:RP}, we mapped five regions in the
$\tilde{a}$-$\tilde{Q}$ phase space, each of which has different phase
behavior. Regions I and II are reminiscent of RN black holes, where there
exist extremal black hole solutions. In these two regions, the global minimum
of the free energy is always at one of the locally stationary points. There is
only one branch of black hole solutions in Region I, while in Region II, there
is a band of temperatures where three branches coexist, and a first-order
LBH/SBH phase transition occurs. In Regions III, IV and V, for low
temperatures, the global minimum of the free energy is at $x=0$, which
describes a naked singularity. For high enough temperatures, the global
minimum is at one of the locally stationary points, and hence there is a
first-order phase transition from the edge state at $x=0$ to the black hole
phase as the temperature increases. In Region III, further increasing the
temperature will lead to another first-order phase transition from Small BH to
Large BH. The phase diagrams with $\tilde{a}=0.05$ and $0.15$ in the
$\tilde{Q}$-$\tilde{T}$ phase space were plotted in FIG. \ref{fig:PT},
respectively. For $\tilde{a}=0.05$, there are three first-order phase
transition lines merging together at a tricritical point. At the critical
temperature, there is a critical point, corresponding to a second-order phase
transition, beyond which there is only one phase. For $\tilde{a}=0.15$, there
is only one phase transition line, which separates the edge state from the
black hole phase. Note that we only focus on spherical topology in our paper,
so it is possible that there are some other states of lower free energy in a
different topological sector with the same charge and temperature. If this
happens, the stable phases discussed above are only metastable.

Using asymptotically AdS boundary conditions, the thermodynamics of BI black
holes was considered in
\cite{IN-Gunasekaran:2012dq,IN-Dehyadegari:2017hvd,IN-Wang:2018xdz}. For the
RN type in Regions I and II, the phase structure of a BI black hole in a
cavity is analogous to that of the corresponding BI-AdS black hole. However
for the Schwarzschild-like type in Regions III, IV and V, we find that there
are some differences between the thermodynamics of the BI black holes under
these two boundary conditions. For example, a LBH/SBH/LBH reentrant phase
transition, which consists of a LBH/SBH first-order phase transition and a
LBH/SBH zeroth-order phase transition, could occur for the BI-AdS black holes
in Region V of \cite{IN-Wang:2018xdz}. On the other hand, such reentrant phase
transition is not observed for any BI black hole in a cavity. Nevertheless, it
is naive to claim that the phase structure of BI black holes depends on
details of the boundary conditions. A\ crucial observation is that, if there
were no edge states, the phase structure of the BI black hole in a cavity
would be quite similar to that of the BI-AdS black hole. In fact, if the edge
state is ignored, the inset of the bottom left panel in FIG. \ref{fig:R45:a}
shows that, as the temperature increases, there is a finite jump in free
energy leading to a zeroth-order phase transition from Large BH to Small BH,
which is followed by a first-order phase transition returning to Large BH.
This LBH/SBH/LBH transition is just the reentrant phase transition.

In asymptotically AdS spaces, the Euclidean action needs to be regulated to
cancel the divergences coming from the asymptotic region. One can use the
background-substraction method to regularize the Euclidean action by
subtracting a contribution from a reference background. The reference
background and the edge state play a similar role in determining the global
minimum of the free energy. For a Schwarzschild-AdS black hole, the reference
background is just the thermal flat space, which is the same as the edge state
at $x=0$ for a Schwarzschild black hole in a cavity. As the temperature
decreases, both the Schwarzschild black hole in a cavity and the
Schwarzschild-AdS black hole thus experience the Hawking-Page transition to
the thermal flat space \cite{IN-York:1986it}. Although there is ambiguity
about the reference background of charged black holes, one can circumvent this
by using the counterterm subtraction method
\cite{CON-Balasubramanian:1999re,CON-Emparan:1999pm}, in which the Euclidean
action is regularized in a background-independent fashion by adding a series
of boundary counterterms to the action. In
\cite{IN-Gunasekaran:2012dq,IN-Wang:2018xdz}, the counterterm subtraction
method was used to compute the Euclidean action for a BI-AdS black hole. For a
RN black hole in a cavity, the global minimum of the free energy is never at
the endpoints, which explains that the phase structures of the RN black hole
in a cavity and the RN-AdS black hole have extensive similarities
\cite{IN-Lundgren:2006kt}. However for a BI black hole in a cavity, there are
some regions in the $\tilde{Q}$, $\tilde{a}$ and $\tilde{T}$ parameter space,
where the global minimum of the free energy is at $x=0$. Different phase
structure from that of a BI-AdS black hole appears there. Our results imply
that, in some region of the parameter space of the BI-AdS black hole, there
might be other states of lower free energy with the same charge and
temperature. Inspired by Schwarzschild-AdS black holes, one natural candidate
is the thermal AdS space filled with charged particles. However, the
backreaction of the charged particles on the AdS geometry should be
considered, which could lead to formation of a naked singularity. It is
inspiring to explore the possible equilibrium phases of lower free energy for
charged AdS black holes.

\begin{acknowledgments}
We are grateful to Zheng Sun and Zhipeng Zhang for useful discussions and
valuable comments. This work is supported in part by NSFC (Grant No. 11005016,
11175039 and 11375121).
\end{acknowledgments}

\appendix

\section{Reduced Action}

In this appendix, we use the reduced action method proposed in
\cite{IN-Braden:1990hw} to evaluate the Euclidean action of a NLED black hole
in a cavity. Instead of employing ansatz functions to solve the equations of
motion of the NLED-gravity system, we here start with the usual form of the
static spherically symmetric metric and consider the reduced action for this
specific metric first. In fact, the Euclidean continuation of the static
spherically symmetric metric takes the form%
\begin{equation}
ds^{2}=b^{2}\left(  r\right)  d\tau^{2}+\alpha^{2}\left(  r\right)
dr^{2}+r^{2}\left(  d\theta^{2}+\sin^{2}\theta d\phi^{2}\right)
,\label{eq:metric}%
\end{equation}
where $b$ and $\alpha$ are free functions of $r$. We suppose that the
Euclidean time is periodic with period $2\pi$. In a canonical ensemble, the
temperature $T$ on the boundary of the cavity at $r=r_{B}$ is fixed, which
impose the boundary condition in terms of the reciprocal temperature:%
\begin{equation}
\int b\left(  r_{B}\right)  d\tau=2\pi b\left(  r_{B}\right)  =T^{-1}.
\end{equation}
At the event horizon at $r=r_{+}$, one has $b\left(  r_{+}\right)  =0$, and
hence the $\tau-r$ part of the metric $\left(  \ref{eq:metric}\right)  $ looks
like a disc. To avoid a conical singularity at $r=r_{+}$, we require that%
\begin{equation}
\alpha^{-1}\left(  r_{+}\right)  b^{\prime}\left(  r_{+}\right)
=1.\label{eq:ab}%
\end{equation}

For the metric $\left(  \ref{eq:metric}\right)  $, the Euclidean continuation
of the action $\left(  \ref{eq:Action}\right)  $ becomes%
\begin{align}
\mathcal{S}^{E}  &  =-16\pi^{2}\int_{r_{+}}^{r_{B}}dr\left\{  \left[  \frac
{1}{\alpha\left(  r\right)  }+\alpha\left(  r\right)  \right]  b\left(
r\right)  +\frac{2rb^{\prime}\left(  r\right)  }{\alpha\left(  r\right)
}\right\}  +32\pi^{2}r_{B}b\left(  r_{B}\right) \nonumber\\
&  -16\pi^{2}r_{+}^{2}-8\pi^{2}\int_{r_{+}}^{r_{B}}drr^{2}a\left(  r\right)
b\left(  r\right)  \mathcal{L}\left(  s,p\right)  . \label{eq:Raction}%
\end{align}
Varying the above action with respect to $\alpha\left(  r\right)  $ and
$b\left(  r\right)  $ gives that%
\begin{equation}
E_{\tau\tau}=\frac{1}{2}T_{\tau\tau}\text{ and }E_{rr}=\frac{1}{2}%
T_{rr}\text{,} \label{eq:REOM}%
\end{equation}
where $T_{\mu\nu}$, the energy-momentum tensor for the NLED field, is given by
$\left(  \ref{eq:stNLED}\right)  $, and $E_{\mu\nu}=R_{\mu\nu}-\frac{1}%
{2}Rg_{\mu\nu}$ is the Einstein tensor. Note that $E_{\tau}^{\tau}$ and
$E_{r}^{r}$ are
\begin{equation}
E_{\tau}^{\tau}=\frac{1}{\alpha^{2}\left(  r\right)  r^{2}}-\frac{1}{r^{2}%
}-\frac{2\alpha^{\prime}\left(  r\right)  }{\alpha^{3}\left(  r\right)
r}\text{ and }E_{r}^{r}=\frac{1}{\alpha^{2}\left(  r\right)  r^{2}}-\frac
{1}{r^{2}}+\frac{2b^{\prime}\left(  r\right)  }{rb\left(  r\right)  \alpha
^{2}\left(  r\right)  }.
\end{equation}

For the NLED field, we take the static spherical symmetry and gauge symmetry
into account and assume that
\begin{equation}
A_{\mu}dx^{\mu}=A_{\tau}\left(  r\right)  d\tau.
\end{equation}
For this ansatz, the equations of motion $\left(  \ref{eq:REOM}\right)  $
becomes%
\begin{align}
\frac{1}{\alpha^{2}r^{2}}-\frac{1}{r^{2}}-\frac{2\alpha^{\prime}}{\alpha^{3}r}
&  =\frac{1}{2}\left[  \mathcal{L}\left(  -\frac{\alpha^{-2}b^{-2}}{2}A_{\tau
}^{\prime2},0\right)  +G^{r\tau}A_{\tau}^{\prime}\right]  \text{,}%
\label{eq:Rtt}\\
\frac{1}{\alpha^{2}r^{2}}-\frac{1}{r^{2}}+\frac{2b^{\prime}}{rb\alpha^{2}} &
=\frac{1}{2}\left[  \mathcal{L}\left(  -\frac{\alpha^{-2}b^{-2}}{2}A_{\tau
}^{\prime2},0\right)  +G^{r\tau}A_{\tau}^{\prime}\right]  \text{,}%
\label{eq:Rrr}%
\end{align}
where $G_{r\tau}=\mathcal{L}^{\left(  1,0\right)  }\left(  -\frac{\alpha
^{-2}b^{-2}}{2}A_{\tau}^{\prime2},0\right)  A_{\tau}^{\prime}$. Moreover,
varying the reduced action $\left(  \ref{eq:Raction}\right)  $ with respect to
$A_{\tau}$ leads to%
\begin{equation}
\partial_{r}\left(  r^{2}G^{r\tau}\right)  =0.\label{eq:RNLED}%
\end{equation}
Substracting eqns. $\left(  \ref{eq:Rtt}\right)  $ from $\left(
\ref{eq:Rrr}\right)  $, one finds that%
\begin{equation}
\frac{\alpha^{\prime}}{\alpha}=-\frac{b^{\prime}}{b}\Rightarrow\alpha\left(
r\right)  =Cb^{-1}\left(  r\right)  ,
\end{equation}
where $C$ is a constant. Actually, $C$ can be determined by eqn. $\left(
\ref{eq:ab}\right)  $:%
\begin{equation}
C=-\frac{\alpha^{3}\left(  r_{+}\right)  }{\alpha^{\prime}\left(
r_{+}\right)  }.
\end{equation}
We can rescale $\tau\rightarrow C\tau$, and hence we have $b\rightarrow
C^{-1}b=\alpha^{-1}$. So after this rescaling, the metric $\left(
\ref{eq:metric}\right)  $ becomes
\begin{equation}
ds^{2}=f\left(  r\right)  d\tau^{2}+\frac{dr^{2}}{f\left(  r\right)  }%
+r^{2}\left(  d\theta^{2}+\sin^{2}\theta d\phi^{2}\right)  ,
\end{equation}
where we define $f\left(  r\right)  =\alpha^{-2}\left(  r\right)  $, and the
period of $\tau$ is $2\pi C$.

Since $A_{\tau}$ and $G^{r\tau}$ is rescaled by $A_{\tau}\rightarrow
C^{-1}A_{\tau}$ and $G^{r\tau}\rightarrow CG^{r\tau}$, respectively, eqns.
$\left(  \ref{eq:Rtt}\right)  $ and $\left(  \ref{eq:RNLED}\right)  $ becomes%
\begin{align}
\frac{f\left(  r\right)  }{r^{2}}-\frac{1}{r^{2}}+\frac{f^{\prime}\left(
r\right)  }{r}  &  =\frac{1}{2}\left[  \mathcal{L}\left(  -\frac{A_{\tau
}^{\prime2}\left(  r\right)  }{2},0\right)  +G^{r\tau}A_{\tau}^{\prime}\left(
r\right)  \right]  ,\\
\partial_{r}\left(  r^{2}G^{r\tau}\right)   &  =0,
\end{align}
which are just the Euclidean version of eqns. $\left(  \ref{eq:ttEOM}\right)
$ and $\left(  \ref{eq:NLEDEOM}\right)  $. Therefore, the solutions to the
above equations are%
\begin{align}
f\left(  r\right)   &  =1-\frac{r_{+}}{r}+\frac{1}{2r}\int_{r_{+}}^{r}%
drr^{2}\mathcal{L}\left(  -\frac{A_{\tau}^{\prime2}\left(  r\right)  }%
{2},0\right)  -\frac{iQ}{2r}A_{\tau}\left(  r\right)  ,\nonumber\\
G^{r\tau}  &  \equiv\mathcal{L}^{\prime}\left(  -\frac{A_{\tau}^{\prime
2}\left(  r\right)  }{2},0\right)  A_{\tau}^{\prime}\left(  r\right)
=-\frac{iQ}{r^{2}}, \label{eq:ESolution}%
\end{align}
where $Q$ is the charge of the black hole as discussed before. Plugging the
solutions $\left(  \ref{eq:ESolution}\right)  $ into the the Euclidean action
$\left(  \ref{eq:Raction}\right)  $, we find that%
\begin{equation}
\mathcal{S}^{E}=\frac{16\pi r_{B}}{T}\left[  1-\sqrt{f\left(  r_{B}\right)
}\right]  -S,
\end{equation}
where $S=16\pi^{2}r_{+}^{2}$ is the entropy of the black hole.

\end{document}